# *Kepler* Science Operations

Michael R. Haas[1,7], Natalie M. Batalha[2], Steve T. Bryson[1], Douglas A. Caldwell[3], Jessie L. Dotson[1], Jennifer Hall[4], Jon M. Jenkins[3], Todd C. Klaus[4], David G. Koch[1], Jeffrey Kolodziejczak[5], Chris Middour[4], Marcie Smith[1], Charles K. Sobeck[1], Jeremy Stober[6], Richard S. Thompson[4], and Jeffrey E. Van Cleve[3]

[1]NASA-Ames Research Center, MS 244-30, Moffett Field, CA 94035
[2]San Jose State University/NASA-Ames Research Center, MS 244-30, Moffett Field, CA 94035
[3]SETI Institute/NASA-Ames Research Center, MS 244-30, Moffett Field, CA 94035
[4]Orbital Sciences Corporation/NASA-Ames Research Center, MS 244-30, Moffett Field, CA 94035
[5]Space Science Office, VP62, NASA-Marshall Space Flight Center, Huntsville, AL 35805
[6]Ball Aerospace and Technology Company, Boulder, CO 80301

[7]E-mail: Michael.R.Haas@nasa.gov

**Abstract.** *Kepler's* primary mission is a search for earth-size exoplanets in the habitable zone of late-type stars using the transit method. To effectively accomplish this mission, *Kepler* orbits the Sun and stares nearly continuously at one field-of-view which was carefully selected to provide an appropriate density of target stars. The data transmission rates, operational cycles, and target management requirements implied by this mission design have been optimized and integrated into a comprehensive plan for science operations. The commissioning phase completed all critical tasks and accomplished all objectives within a week of the pre-launch plan. Since starting science, the nominal data collection timeline has been interrupted by two safemode events, several losses of fine point, and some small pointing adjustments. The most important anomalies are understood and mitigated, so *Kepler's* technical performance metrics have improved significantly over this period and the prognosis for mission success is excellent. The *Kepler* data archive is established and hosting data for the science team, guest observers, and public. The first data sets to become publicly available include the monthly full-frame images, dropped targets, and individual sources as they are published. Data are released through the archive on a quarterly basis; the *Kepler Results Catalog* will be released annually starting in 2011.

Subject headings: space vehicles — instruments: telescopes

Submitted to: *Astrophysical Journal Letters*

## 1.0 INTRODUCTION

The *Kepler* spacecraft was launched into a 372.5-day, earth-trailing, heliocentric orbit from Cape Canaveral Air Force Station aboard a Delta II 7925-10L on March 7, 2009 (UTC). In this orbit, the photometer must be rolled 90° about its axis every 93 days to keep the solar arrays illuminated and the focal-plane radiator pointed away from the Sun. The nominal roll dates shown in Table 1 determine the primary cycle for science operations.

Data collection is interrupted once a month to earth-point the high-gain antenna and download the stored science and engineering data. A download takes about 6 hours over Ka-band. The duration of the breaks is significantly longer because calibration data are collected before and after each roll, and the photometer must thermally re-equilibrate before returning to fine point. The on-board solid state recorder (SSR) holds approximately two months data, so data which are unsuccessfully downloaded one month are saved and downloaded during the next monthly contact. To date, the Ka-band performance has been excellent with less than 0.1% of the data requiring retransmission.

*Kepler* can simultaneously observe up to 170,000 targets at its long-cadence (LC) sampling interval of 29.4 min and up to 512 targets at its short-cadence (SC) sampling interval of 58.8 sec. Every 48 LCs, about once per day, a small subset of the targets are extracted from one LC and stored separately as reference pixels (RP). These RP and a small amount of engineering data are downloaded biweekly via X-band to monitor spacecraft health and performance. No data breaks are introduced because the requisite low-gain antenna is omni-directional.

Every 3.0 days, one or more reaction wheels approach their maximum operating angular velocity as they absorb angular momentum due to solar radiation torques on the spacecraft. In order to "desaturate" the wheels, the spacecraft is transitioned from fine point, hydrazine thrusters are fired to dump the excess momentum, and the spacecraft is commanded back to fine point a few minutes later. Each "desat" causes significant degradation of the photometric precision for one LC and several SC intervals.

## 2.0 COMMISSIONING ACTIVITIES

The commissioning activities began immediately after launch and lasted 67 days as summarized in Table 2. Before dust cover eject (DCE), dark data was collected at several different temperatures for use in photometric calibration; both forward-clocked full-frame images (FFIs) and reverse-clocked long-cadences (RC LCs) were obtained. After recovering from a safe-mode event on Day-of-Year (DOY) 83 and conducting a series of reviews, DCE occurred on DOY 98. The first images showed that the pre-launch focus met the encircled-energy requirements. However, the LC target set is pixel-limited, so small piston and tip/tilt adjustments were used to sharpen focus and minimize target loss. A mean 95% encircled-energy diameter of 4.3 pixels was achieved. After the focal-plane geometry (FPG) was measured, target tables were defined for use in mapping the pixel response function (PRF; Bryson *et al*. 2010). While the FPG and PRF data were being processed and the Quarter 1 (Q1) target tables were being prepared, eight FFIs were collected at convenient times over several days to estimate stellar variability. The last major commissioning activity was 9.7 days of cadence data collection (referred to as "Quarter 0" or Q0) to estimate the combined differential photometric performance (CDPP; Jenkins *et al*. 2010a) and identify unclassified or misclassified stars. Most relatively isolated stars in the *Kepler* field-of-view (FOV) with magnitudes brighter than 13.6 were observed, independent of spectral type.

## 3.0 TARGET TABLES

A full-frame image (FFI) is collected each month to confirm the target apertures, monitor photometer health and performance, and provide important calibration data. These FFIs include all 96.5 million pixels from the 42 CCDs (x 2 amplifiers/CCD = 84 channels). Each science pixel views 3.98 x 3.98 arcsecs on the sky, so the active FOV is 115.6 sq degrees. In contrast, routine data collection saves and downloads less than 6% of the pixels to minimize on-board processing and storage and to limit the duration of download breaks.

To collect the desired pixels, an aperture is defined for each target star using an entry in a target table. New LC, SC, and RP target tables are required every quarter because the 90° rolls move the stars to new positions in the focal plane. One LC target table is flown each quarter, but the SC target table is changed monthly to provide greater flexibility in target selection for this limited, but valuable resource. The focal plane was designed to be as four-fold symmetric as possible in order to maximize the number of LC targets that can be observed continuously throughout the mission. A small central asymmetry and ±3 pixel errors in the CCD positions force 1% of the target stars to vary by quarter.

An extensive stellar classification program measured magnitudes, effective temperatures, and surface gravities for millions of stars in the *Kepler* FOV. The resulting *Kepler Input Catalog* (http://archive.stsci.edu/kepler/kepler_fov/search.php) was used to identify those stars for which transits by earth-size planets in the habitable zone are most readily detectable (Batalha *et al*. 2010a). The target set also includes stars for asteroseismology (Gilliland *et al*. 2010), astrometry (Monet *et al*. 2010), cluster studies, a control group, and the guest observer program. Each quarter these various target lists are individually prioritized, assigned pixel allocations, and combined into a master target list. For each target, an optimal aperture is defined that includes all pixels that improve the signal-to-noise ratio (SNR) for simple aperture photometry. This calculation uses the measured pixel response function (PRF; Bryson *et al*. 2010), on-orbit noise characteristics of the focal plane (Caldwell *et al*. 2010), *Kepler* magnitude (Koch *et al*. 2010), and contamination by nearby stars. The optimal apertures are enlarged to account for differential velocity aberration, which varies annually by 1.5 pixels, and an additional halo of pixels is added to accommodate pointing errors, changes in focus, *etc*. Finally, one of the 1024 unique aperture masks is assigned to efficiently capture the specified pixels. Since the individual-pixel data are downloaded, subsequent photometric processing can include those pixels which maximize the SNR.

The LC and SC target tables are limited to 5.44 million and 43,520 pixels, respectively, by the *Kepler* design. Consequently, the Q0 data collected during commissioning was limited to 52,496 LC targets and 310 SC targets because double halos were used to compensate for early uncertainties in focal-plane geometry and PRF (Bryson *et al*. 2010) and because this target set included many bright stars, which require more pixels on average. During science operations, the number of LC targets remains pixel-limited, but has increased from 156,097 in Q1 to 165,716 in Q3 due to improvements in defining and assigning aperture masks. The number of SC targets has been 512 for every month except month 2 of Q2, which was limited to 479 targets because brighter stars were observed.

The RP target table is limited to 96000 pixels, or about 320 targets. The RP targets are distributed over all 84 channels, but their density is larger on the focal-plane periphery to increase the precision of the attitude determination. RP aperture definitions require more pixels per target than cadence targets because they have double halos and collect their own collateral data and background pixels.

## 4.0 DATA COLLECTION, FLOW, AND PROCESSING

New target tables are generated and uploaded a few weeks before the end of each quarter in anticipation of the upcoming roll. Each month's data collection begins with 3 reverse-clock (RC) LCs for calibration, includes about 1500 LCs and 45000 SCs, and ends with 1 FFI and 3 RC LCs. These data and the associated engineering data are downloaded from the flight segment using the Deep Space Network (DSN) stations at Canberra, Madrid, and Goldstone. The telemetry stream is packetized and encoded on-board the spacecraft, transmitted to earth via Ka-band at 4.33 Mbps, unencoded by the DSN, and unpacked by the Mission Operations Center (MOC) at the Laboratory for Atmospheric and Space Physics (LASP) in Boulder, CO. Ball Aerospace & Technology Company (BATC) is responsible for *Kepler* Mission Operations, providing engineering support and directing the *Kepler* operations team at LASP. From the MOC, the data flows to the Data Management Center (DMC) at the Space Telescope Science Institute (STScI) in Baltimore, MD, where the science data are uncompressed, correlated with the appropriate engineering data, packaged into Flexible Image Transport System (FITS) files, and sent to the Science Operations Center (SOC) at the NASA-Ames Research Center, Moffett Field, CA.

The SOC is responsible for calibrating the raw-pixel data, rejecting cosmic rays, generating light curves via aperture photometry, detrending against relevant engineering data, searching for threshold-crossing events (TCEs), and validating these TCEs to distinguish possible planetary candidates from non-astrophysical false positives (Jenkins *et al*. 2010b). The processed data are evaluated and approved by the Data Analysis Working Group (DAWG), which is comprised of SOC, Science Office, and Science Team members, and then sent to the DMC for distribution through the Multi-mission Archive at STScI (MAST). The processed data are also provided to the TCE Review Team (Batalha *et al*. 2010b), who evaluate and prioritize the TCEs for ground-based follow-up observations (Gautier *et al*. 2010).

RPs are downloaded from the flight segment biweekly via X-band at rates decreasing from 8 Kbps to less than 1 Kbps as *Kepler* drifts away from the Earth. These data flow directly from the MOC to the SOC, where they are processed and used to monitor pointing, brightness, plate scale, CDPP, and other photometer health metrics (Caldwell *et al*. 2010). RPs are also collected after each quarterly roll and each monthly download to verify the science attitude (i.e., RA = $19^h22^m40^s$ and Dec = 44°30'00") and perform small pointing adjustments as required. Additional performance metrics are computed on a monthly basis using 200 unsaturated, relatively isolated stars per channel once the LC science data are downloaded.

## 5.0 SCIENCE MISSION TIMELINE

Science operations began on May 13, 2009 (UTC). Figure 1 compares the nominal science mission timeline with the on-orbit performance. Safemode events occurred on DOY 166 and 183 with nearly identical signatures. Both events turned the photometer off, which created several-day gaps in the cadence data and introduced significant thermal transients for the first few days after the resumption of science. A root cause investigation identified two radiation-susceptible components in the microprocessor power-on reset circuitry and confirmed them with radiation testing. In order to minimize the impact of similar events in the future, the safemode-recovery procedure has been modified to keep the photometer powered-on following such commanded resets.

The spacecraft experienced a loss of fine point (LOFP) on DOY 224, 256, 268, and 282 for periods of 4 to 22 hrs, but fine-point was readily restored once commanded. These anomalies appear to involve star-tracker interactions with the fine-guidance-sensors. Root cause is identified and fixed for the first LOFP. Root cause for the others is under investigation, but mitigating actions have been identified and deployed.

Figure 1 shows that the science data collection prior to the first quarterly roll (DOY 169) was shortened because commissioning ran long. To compensate, the monthly download on DOY 141 was canceled and a single SC target table was used for Q1. Data collection was terminated after 33.5 days by the safemode on DOY 166. Note that Q0 and Q1 have the same roll angle, but used substantially different target tables. The first full observing quarter is Q2, which started and ended on schedule. Since data was downloaded while recovering from the DOY-183 safemode, the mid-July monthly was canceled, but the SC target table was changed per plan on DOY 201. The August monthly was nominal and Q3 has been nominal to date, except for the two LOFP indicated.

The nominal timeline has science breaks of 42 hours at the quarterly rolls and 26 hours for the intervening monthly downloads. The quarterly breaks require extra time because a small post-roll pointing adjustment is required to achieve the precise science attitude assumed when generating the target apertures. The quarterly science breaks are expected to decrease in duration after the first year when all roll angles are calibrated. The achieved data completeness, including the safemodes and desaturations, is 92%, which is consistent with the pre-launch plan.

## 6.0 PERFORMANCE METRICS

To determine pointing drift, the daily RP data are used to measure the RA, Dec, and roll errors relative to the nominal science attitude. These quantities predict the maximum deviation of any pixel from its nominal location assuming rigid-body motion (i.e., no change in plate scale or other distortion of the focal plane). This metric, known as the maximum attitude residual (MAR), is plotted in Figure 2. Q1 and Q2 show significant pointing drift, which was controlled by performing small pointing adjustments to protect the integrity of the target aperture definitions. After Q2, the fine-guidance-sensor pixel-of-interest blocks were individually centered on their stars and a 3-DN electronic bias was removed as unwanted background signal. These changes reduced the pointing drift (i.e., MAR) to a few mpix per month in Q3. The root-sum-square drift on time scales of hours to days (i.e., the observed pointing detrended against a 5-day moving median of the pointing) decreased from 1.25 mpix in month 3 of Q2 to 0.22 mpix in month 1 of Q3. Another significant change was the removal of an eclipsing binary star with a 1.7-day period and a variable star with a 3.3-day period from the ensemble of 40 fine-guidance stars. Though not evident in Figure 2, the signature of both stars is apparent in the raw light curves and centroid data for Q2 (Jenkins *et al.* 2010a).

*Kepler* data is highly compressed to minimize on-board storage requirements and data download times. Both are important in minimizing the frequency and duration of the required science breaks. Data compression involves requantization of the analog-to-digital converted values, baseline subtraction, and Huffman encoding. In order to requantize all CCD channels and both cadences with one requantizaton table, a channel-specific mean black is subtracted from each data value and a common fixed offset is added to eliminate negative numbers. Requantization reduces the original 23-bit pixel data values to 16 bits, while adding no more than one-fourth of the intrinsic measurement uncertainty. These requantized values are differenced from a daily baseline cadence before Huffman encoding. Figure 3 shows the LC data compression for Q1 and the first month of Q3. The minima, which correspond to maximum data compression, occur immediately after the daily baseline images. The gradual increase in bits/pixel over each day-long period results from cosmic ray strikes, pointing drift, stellar variability, eclipsing binaries, transiting planets, and other phenomena which further change each subsequent image. During Q3, the compression is poor on DOY 262 because of thermal transits associated with the just-completed monthly download and on DOY 266 because of a small, spurious increase in background during the baseline image (i.e., 'Argabrightening,' Van Cleve & Caldwell, 2009). The gaps beginning on DOY 268 and 282 are due to LOFP. Excluding these events, the variations in compression decreased from Q1 to Q3 and the mean compression improved from 4.8 to 4.5 bits/pix as a result of improved pointing (cf Figure 2).

## 7.0 ARCHIVAL PRODUCTS

The *Kepler Input Catalog (KIC),* which contains roughly 4.5 million stars on *Kepler* silicon and 13 million stars overall, was recently released to the public and is available for guest-observer target selection. The *Kepler Results Catalog (KRC)* will compile the results of the planetary search effort and the associated ancillary investigations. Stellar parameters such as parallax, age, radius, and metallicity will be determined and reported for many stars, whether they have transiting planets or not. The *KRC* is scheduled for release on an annual basis starting in mid-2011, with a final delivery one-year after science data collection ends.

The activities for each quarter begin with final target selection about two months prior to the three-month observing period and end with the data being hosted at MAST (http://archive.stsci.edu/kepler) four months after the last monthly download. This nine-month cycle is repeated for each observing quarter. The data products hosted by MAST include the raw and calibrated target-level pixel data and the associated flux time series. The latter contain light curves and row/column centroids as a function of cadence. Each target for each period has well-defined proprietary periods for access by the Science Team, guest observers, and public. The associated engineering data, calibration files, and focal-plane characterization models used in SOC post-processing are hosted to support archival research. After all proprietary rights have expired for a given quarter, the raw and calibrated pixel data will be made available by module, not just by target.

Some *Kepler* data are scheduled for rapid public release. The monthly FFIs are calibrated and delivered to MAST for public release four months after the observing period ends. The data from planetary-search targets are released 60 days after the *Kepler* Science Team formally drops them. These dropped targets include stars that were originally classified as dwarfs, but found to be giants (Koch *et al*. 2010), stars that are too variable for transit detection, and "stars" that are identified as KIC artifacts. The 8441 targets dropped before Q2 are available at MAST. *Kepler* anticipates dropping thousands of additional targets as the telemetry rates decrease.

Planetary-search data are scheduled for public release as follows: Q0 and Q1 in June, 2010; Q2 in June, 2011; Q3/Q4 in June, 2012; Q5/Q6 in June, 2013; and the remainder at end of mission in November, 2013. Data collected by the science team for ancillary studies are subject to a one-year proprietary period which starts upon MAST ingest. The proprietary period for each guest-observer cycle is 12 months from MAST ingest or 6 months after the last data are ingested, whichever comes later. The first guest-observer cycle commenced observations in June, 2009. Targets that are exclusively for engineering purposes have no proprietary period; most of these targets are collections of pixels required for calibration or instrument-health, but a few astrophysical sources may be included as the mission proceeds. The data on individual sources are publically released once published in a scientific journal.

Periodic reprocessing will occur as the data is better understood and the SOC pipeline matures. A new pipeline release is scheduled about once a year and will trigger a complete reprocessing of all science data. Each official data release is accompanied by detailed release notes and identified by populating the keywords DATA_REL and QUARTER in each FITS file. Additional information can be found in the *Kepler Instrument Handbook* (Van Cleve & Caldwell, 2009), *Kepler Archive Manual*, and *Kepler Data Analysis Handbook* at http://archive.stsci.edu/kepler.

## 8.0 CONCLUSIONS

*Kepler* has successfully completed its commissioning phase, entered science operations, and collected 8 months of science data. Some data are already available through the public archive. Given *Kepler's* excellent on-orbit performance and 10-yr supply of expendables, an extension beyond the nominal 3.5-yr mission is worthy of serious consideration. Funding for this Discovery mission is provided by NASA's Science Mission Directorate.

*Facilities: Kepler.*

## FIGURES

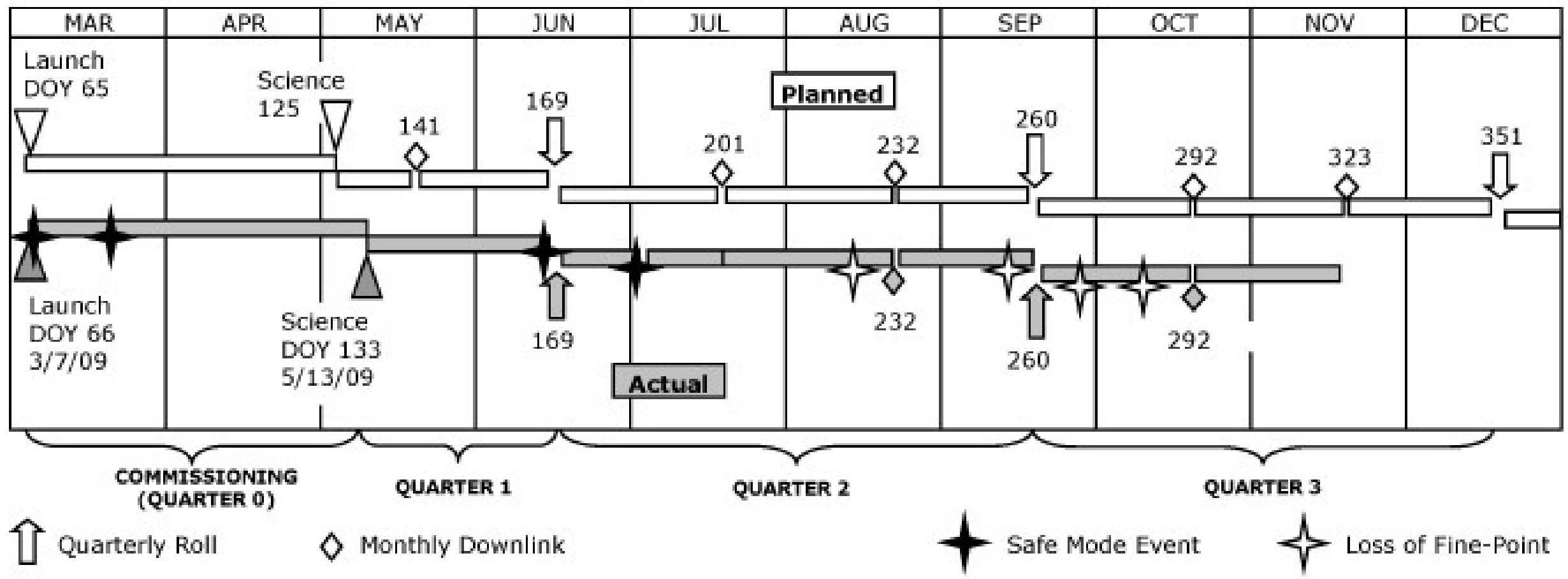

Figure 1. The planned science mission timeline is compared with the actual timeline from launch through the end of 2009. Dates and Day-of-Year (DOY) are UTC.

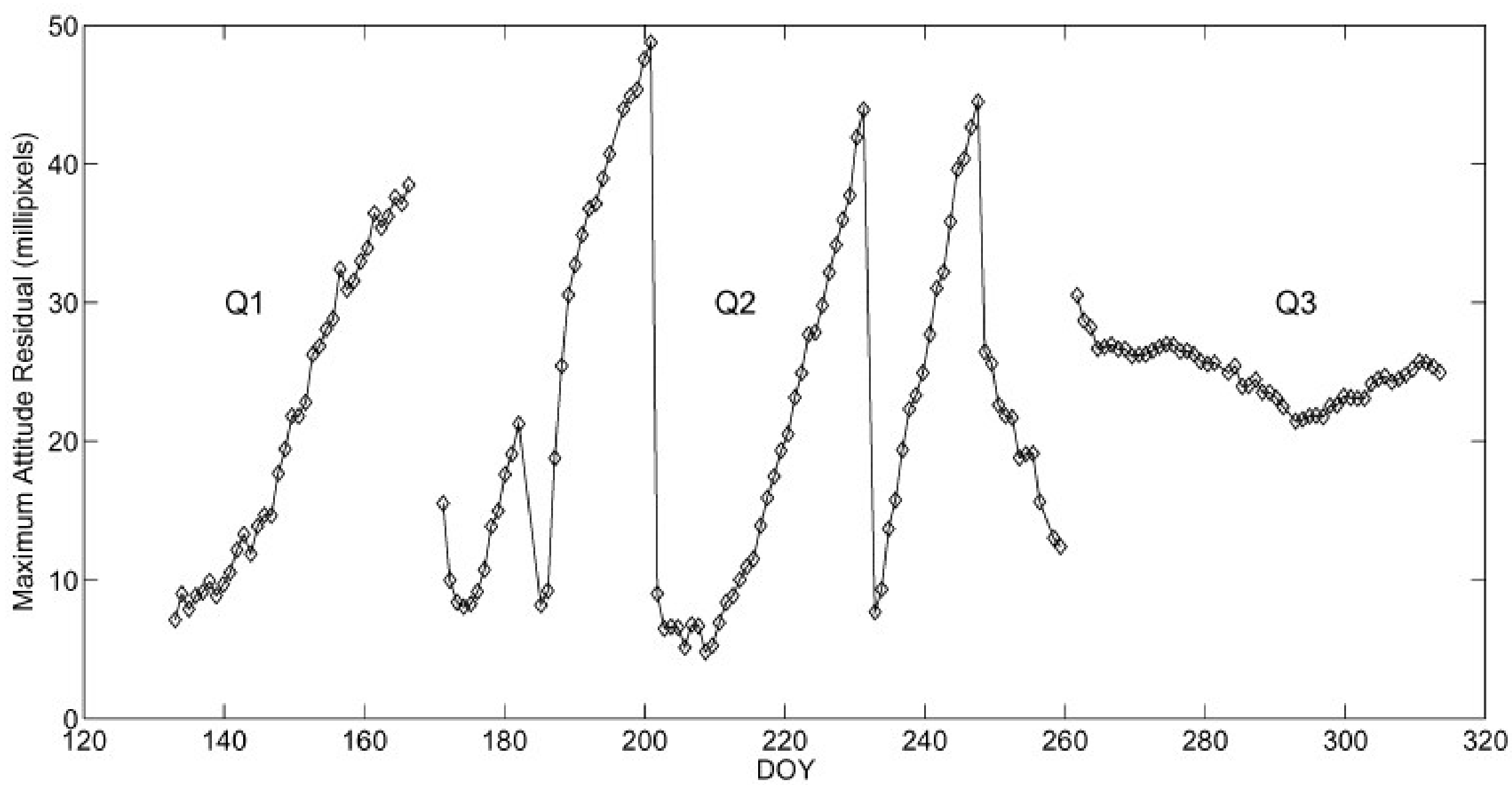

Figure 2. The maximum attitude residual, a sensitive measure of *Kepler's* pointing drift, is plotted for the period spanning Quarters 1, 2, and part of 3. Safe mode events occurred on day-of-year 166 and 183, and pointing adjustments occurred on DOY 201, 232, and 247. The last pointing adjustment overcompensated, so MAR decreased for the last 12 days of Q2.

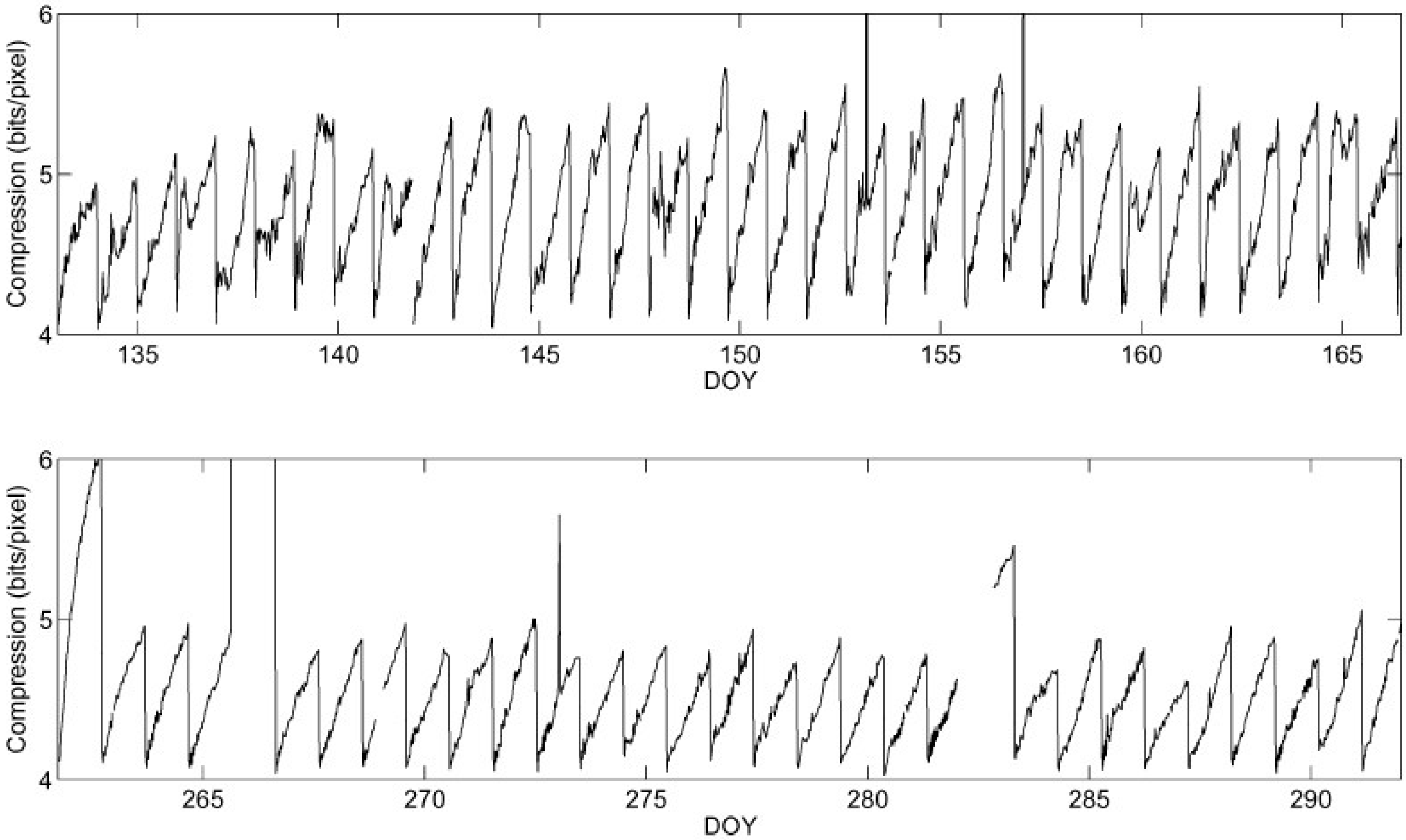

Figure 3. The mean data compression efficiency for all 84 channels as a function of Day-of-Year for Quarter 1 (upper panel) and month 1 of Quarter 3 (lower panel).

| Table 1. Quarterly Roll Dates | | | | |
|---|---|---|---|---|
| **Year** | **Spring** | **Summer** | **Autumn** | **Winter** |
| 2009 | --- | 6/18 | 9/17 | 12/17 |
| 2010 | 3/19 | 6/23 | 9/23 | 12/22 |
| 2011 | 3/24 | 6/27 | 9/29 | 12/29 |
| 2012 | 3/29 | 6/28 | 10/1 | --- |

| Table 2. Commissioning Timeline | |
|---|---|
| **DOY (UTC)** | **Activity** |
| 66 | Launch, orbit injection, separation, detumble |
| 67 | Transition to standby; initialize solid state recorder |
| 67, 68, 70 | Check-out x-band over multiple stations |
| 68 | Calibrate coarse sun sensor |
| 68 - 70 | Initialize photometer |
| 70 - 71 | Calibrate high-gain antenna |
| 70 - 80 | Collect dark FFIs and RC LCs |
| 72, 93, 109 | Calibrate Ka-band over multiple stations |
| 75, 79 | Measure cosmic ray distribution |
| 83 - 94 | Enter safe mode and recover |
| 95 | Enable sun avoidance |
| 98 | Eject dust cover |
| 98 | Collect first light image |
| 98 - 100 | Calibrate fine-guidance sensors |
| 99 - 101 | Collect calibration data during cool-down |
| 103 - 112 | Optimize encircled energy (focus) |
| 112 | Determine focal plane geometry |
| 112 - 114 | Measure scattered light and ghosting |
| 114 - 116 | Measure stellar variability (golden FFIs) |
| 116 - 119 | Map pixel response function |
| 120 - 121 | Measure gain and linearity |
| 121 - 131 | Measure combined differential photometric performance (CDPP; Q0) |
| 132 | Adjust science attitude |
| 133 | Start science data collection (Q1) |